\documentclass[12pt,preprint]{aastex631}

\usepackage{graphicx}
\usepackage{amsmath,xcolor,url}
\usepackage{multirow}
\usepackage{amssymb}
\usepackage{epsfig}
\usepackage{hyperref}
\hypersetup{
colorlinks=true,
linkcolor=blue,
filecolor=red,
citecolor=blue,
urlcolor=cyan,
}
\usepackage{natbib}
\begin{document}

\newcommand{\Msol}{\mbox{$M_{\odot}$}}
\newcommand\brobor{\smash[b]{\raisebox{0.6\height}{\scalebox{0.5}{\tiny(}}{\mkern-1.5mu\scriptstyle-\mkern-1.5mu}\raisebox{0.6\height}{\scalebox{0.5}{\tiny)}}}}

\def\kmss {km s$^{-1}$}
\def\kms {km s$^{-1}$}


\title[Shaping Outflows and Jets]{Shaping Outflows and Jets by Ambient Pressure: a Unified Framework}

\author[0000-0003-2171-5074]{Willem A. Baan}
\affiliation{Xinjiang Astronomical Observatory, CAS, 150 Science 1-Street, Urumqi, Xinjiang 830011, P.R. China}
\affiliation{Netherlands Institute for Radio Astronomy ASTRON, NL-7991 PD Dwingeloo, the Netherlands}
\email{baan@xao.ac.cn, baan@astron.nl}

\author[0000-0003-4341-0029]{Tao An}
\affiliation{Shanghai Astronomical Observatory, CAS, 80 Nandan Road, Shanghai 200030, P.R. China}
\affiliation{Xinjiang Astronomical Observatory, CAS, 150 Science 1-Street, Urumqi, Xinjiang 830011, P.R. China}
\affiliation{Guizhou Radio Astronomical Observatory, Guizhou University, 550000, Guiyang, China}
\email{antao@shao.ac.cn}

\begin{abstract}
Astrophysical outflows are ubiquitous across cosmic scales, from stellar to galactic systems. 
While diverse launching mechanisms have been proposed, we demonstrate that these outflows share a fundamental commonality: their morphology follows the physics of pressure-confined supersonic flows. 
By extending classical deLaval nozzle theory to account for ambient pressure gradients, we present a unified framework that successfully describes outflows from young stellar objects to active galactic nuclei. 
This simplified approach compared to full magneto-hydrodynamic treatments, captures the essential physics governing outflow shapes across different scales.
Our model reveals a remarkable consistency in pressure profiles, characterized by a power-law exponent near `minus two' across six orders of magnitude in spatial scale, independent of the internal characteristics of the outflow or the nature of central engine. 
This discovery suggests a universal mechanism for outflow collimation and acceleration, bridging the gap between theoretical models and observational features across a wide range of astronomical scales.
\end{abstract}


\section{Introduction}
\label{sec:intro}

Jets and outflows are widespread phenomena in astrophysical systems operating across a vast range of scales from young stellar objects (YSOs) to active galactic nuclei (AGN) \citep{1999PhR...311..225L, 2002ApJ...568..726S, 2019ARA&A..57..467B}. 
These collimated flows represent a significant channel for momentum and energy transfer between the central compact object and the ambient medium, profoundly impacting the growth of stars and galaxies as well as the cosmic evolution of interstellar and intergalactic environments. 
Despite remarkable progress in high-resolution observations and theoretical modeling over the past decades, the key physical mechanisms governing the launching, acceleration, and collimation of astrophysical outflows remain partially unexplored, especially from an observational perspective \citep{1997ASPC..121..845L, 2007prpl.conf..277P, 2012SSRv..169...27P, 2019ARA&A..57..467B}. 

The absence of a unifying model for the launching and shaping of astrophysical outflows  has led to various theoretical proposals \citep{1974MNRAS.169..395B, 1991ApJ...370L..31S,1999PhR...311..225L, 2007prpl.conf..277P, Marti2019}.
While empirical modeling of bipolar Planetary Nebulae outflows has yielded valuable kinematic parameters,  these models do not yet incorporate the physical processes that shape the outflows \citep{2015A&A...582A..60C,2016A&A...593A..92B}.
For powerful extragalactic sources, leading launching models such as the magneto-centrifugal acceleration \citep{1982MNRAS.199..883B, 2007prpl.conf..277P}, magnetic tower flows \citep{1995ApJ...439L..39U, 1996MNRAS.279..389L}, and black hole extraction models \citep{1977MNRAS.179..433B} invoke complex magnetic and hydrodynamic processes near the central engine,
but these do not fully explain the observed shape and observed emission structure of the jets.
Conversely, relativistic fluid models that incorporate the viscous interaction with the ambient medium can reproduce the observed shape and emission characteristics but do not yet describe the launching conditions of the jets \citep{1980ApJ...239..433B,1984A&A...133..205N,2008ApJ...681...84A, MartiEA2016}.
Direct imaging of the compact launching zones of these AGN sources would contribute significantly to the creation of a unified model.
However, this is near or beyond the limits of current astronomical facilities, except for the nearest AGN jets in M\,87 and Centaurus\, A \citep{2018ApJ...855..128W,2021NatAs...5.1017J}. 
Consequently, most numerical simulations tracking the evolution and collimation of the outflow over long distances necessitate assumptions about the properties of the ambient medium and the jet-environment interactions, whose validity remains unclear in part (see review by \citet{Marti2019}).
As a result, these limitations particularly affect our understanding of the potential influence of the boundary conditions surrounding the central engine on the propagation of the jet over long distances.

Recognizing the potential influence of ambient conditions on the shaping and evolution of outflows, we independently consider the fluid dynamics of a supersonic deLaval nozzle flow as a basis for a phenomenological jet collimation scenario. 
This approach draws inspiration from engineering applications of the deLaval nozzle theory in thrusters for rockets and supersonic aircraft, where the fluid first experiences subsonic acceleration in a narrowing cross section until it reaches the sound speed at the throat of the nozzle (Mach number = 1) followed by supersonic acceleration forced by a widening cross section towards the exit of the nozzle.
This model essentially encodes a pressure gradient on the rigid walls of the nozzle, which continuously mediates the acceleration and expansion of the flow by converting thermal energy into kinetic energy.
 
Our key introduction lies in adapting the deLaval nozzle equations to astrophysical environments by replacing the confining nozzle walls with an ambient pressure declining with distance.  
In this manner, the ambient pressure gradient provides the pressure balance in the boundary and determines the variation of the Mach number, the velocity, and the shape of the outflow as a function of distance.
By comparing the analytical shape of the supersonic part of the outflow model with observed astrophysical outflows, we can determine the best-fit values for pressure gradient and the scale length of the gradient, as well as the location and size of the throat that define the launching point of the outflow.
While the importance of pressure gradients in shaping outflows was recognized early by \citet{1980ApJ...239..433B} and explored further by \citet{MartiEA2016}, a comprehensive analytical treatment incorporating pressure gradients with gravitational effects has not been fully implemented in recent theoretical studies.

A possible complication in modeling astrophysical outflows is that they originate close to compact central engines and experience gravitational deceleration, particularly at the initial expansion stage. 
To account for gravity effects, a term expressing the differential deceleration in the velocity may be added to the local Mach number (see Section \ref{sec:methods}).
A sustained and non-stagnated outflow from the nozzle is then only possible when the sound speed at the launching point (the nozzle throat) is significantly larger than the local gravitational escape velocity, which requires a minimum threshold value for the effective local temperature.
Gravitational deceleration reduces the size and distance of the throat and causes a faster initial expansion of the flow close to the source, while the shape of the rest of the outflow remains unchanged.
Gravitational deceleration may explain the proximity of the launching point and the larger opening angle of the outflows observed in compact Galactic stellar sources and supermassive extragalactic sources.

\section{Methods} \label{sec:methods}

\subsection{A  deLaval Nozzle Flow Model}

Our novel approach to modeling astrophysical outflows adapts the equations governing the ideal gas flow through a deLaval nozzle by incorporating an ambient pressure gradient environment. 
In the classical deLaval nozzle as described above, the fluid flows through two regimes: (a) subsonic acceleration within a narrowing channel as the walls constrict until velocity reaches the sound speed at the throat (Mach = 1); (b) supersonic acceleration of the flow forced by the diverging walls towards the nozzle exit.
For an idealized case of incompressible flow between fixed nozzle walls only adopting adiabatic behavior and continuity of the flow, the variation of the shape of the wall determines the variation of the pressure and all other flow parameters by setting the Mach number at each position. 
Following established fluid dynamics principles  \citep{1987flme.book.....L, 1998pfp..book.....C}, the relationship between Mach number and nozzle area is expressed as::
\begin{equation}
\frac{A}{A^*} = \frac{1}{M} \left[ \frac{1 + {\frac{\gamma-1}{2}}M^2}{\frac{\gamma + 1}{2}} \right]^{(\gamma +1)/2(\gamma -1)} 
\end{equation}




\noindent where $A^*$ is the throat area at the sonic point ($M=1$), with the Mach number $M$ defined as the ratio of the local flow velocity $V$ to the local speed of sound $c_s$: $M = {V}/{c_{\rm s}}$.
The speed of sound $c_{\rm s}$ depends on the ratio of the local gas temperature $T$ and the local mass density $\rho$. 

\begin{figure*}
\centering
\includegraphics[width=0.8\textwidth]{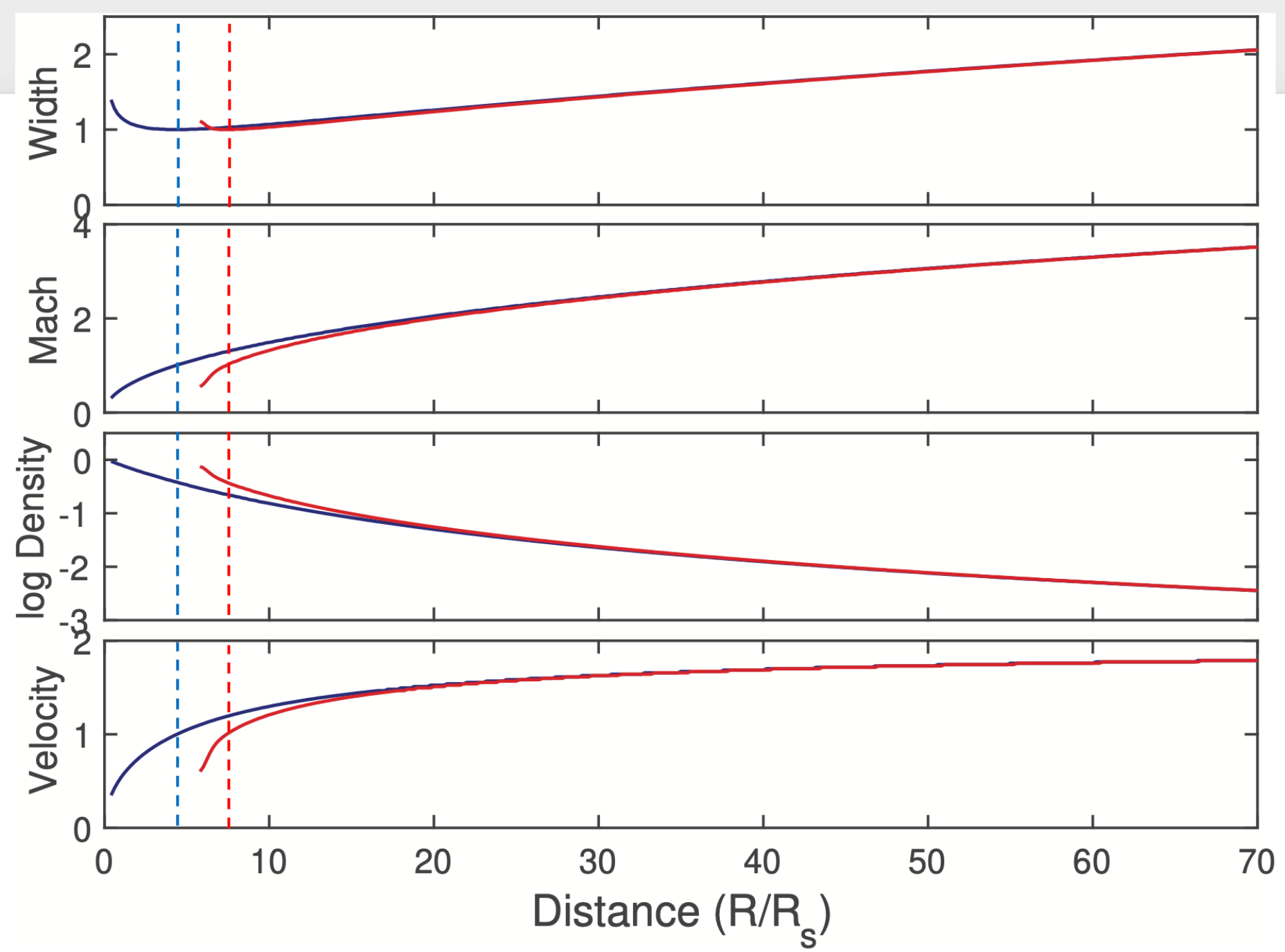}
\caption{The parameters of the outflow as it passes from subsonic to supersonic across the nozzle launching point at Mach 1. 
The pressure gradient has an index $\alpha = -2$, and the X-axis is the travel distance in units of the pressure gradient scale length.
 The Y-axis labels are defined as: (top frame) the Width  normalized by the width of the nozzle at the sonic point, (second frame) the Mach number,
 (third frame) the logarithm of density $\rho$ normalized with the density $\rho_0$ at the nozzle entry, and (bottom frame)is the flow velocity normalized by the sound speed at the sonic point.
As the outside pressure decreases, the outflow width after the throat increases, along with the Mach number and flow velocity, while the density and the temperature decrease.  
The blue curves represent an outflow without gravitational deceleration, and the red curves include a nominal degree of deceleration. 
The red curves show the rapid and delayed increase in the velocity and the Mach number as the launch point of the outflow moves to a lower ambient pressure because of the gravitational deceleration. 
The blue and red dotted lines indicate the location of the Mach = 1 nozzle point in the flow, which is the transition from subsonic to supersonic flow.  
For both cases, the outflow velocity along the displayed section increases by a factor of two, while the Mach number increases by a factor of 3.5 along the displayed distance.}
\label{fig:Mfig1}
\end{figure*}

The value of the funnel area $A$ in the equation at position $R$ above determines the local value of $M$ in the flow, which then determines the other key flow parameters at position $R$ for a nozzle flow as:
\begin{align}
\frac{T}{T_0} &= \left[1 + \frac{\gamma-1}{2}M^2\right]^{-1}, \frac{\rho}{\rho_0} &= \left[1 + \frac{\gamma-1}{2}M^2\right]^{-1/(\gamma-1)}, \frac{P}{P_0} &= \left[1 + \frac{\gamma-1}{2}M^2\right]^{-\gamma/(\gamma-1)},
\end{align}
\noindent where the subscript 0 denotes the initial quantities at the subsonic nozzle inlet. 
Equations (1) and (2) describe the flow properties inside a deLaval nozzle with a rigid boundary. 

In order to adapt this model to astrophysical outflows, we replace the rigid nozzle walls with a surrounding ambient pressure gradient distribution as a function of traveling distance $R$ given by:
\begin{equation}
P(R) = P_0\left(\frac{R}{R_{\rm s}}\right)^{\alpha} + P_2,
\end{equation}
where $R_s$ is the scale length over which the pressure declines and $\alpha$ is power law exponent with a negative value. 
$P_0$ is a reference pressure taken  at the inlet of the nozzle. 
The additional term $P_2$ accounts for an additional pressure component reflecting a much lower and constant pressure of the ambient medium away from the source as well as the development of the boundary layer. 
The external pressure confines the outflow boundary in a manner similar to those rigid deLaval nozzle walls. 

Balancing the internal and external pressures, the local Mach number now follows from the varying pressure (from eq. 2) and from the travel distance $R$ as:
\begin{equation}
M(R)^2 = 
\frac{2}{\gamma-1}\left[\left(\frac{P(R)}{P_0}\right)^{1-\gamma/\gamma} - 1\right] = 
\frac{2}{\gamma-1}\left[\left(\frac{R}{R_s}\right)^{\alpha(1-\gamma)/\gamma} -1 +P_2^{(1-\gamma)/\gamma}\right] ,
\end{equation}
\noindent resulting in an increasing Mach number for a negative value for $\alpha$.
Given the value for $M(R)$, the local width of the flow (from eq.1), and the density and temperature (from eq.2) can be determined as a function of travel distance $R$. 
The velocity in the flow follows from the continuity equation as $V(R)$ = $1/\sqrt{\rho(R) A(R)}$.

The above Eq. (4) also determines the location of the Mach One nozzle, which is the launching point of the supersonic jet.
When using $\alpha = -2$ and an adiabatic constant $\gamma$ = 5/3, as used in the simulations in the next section, and ignoring $P_2$ close to the source, the subsonic flow only needs a distance of $R$ = 1.43 $R_s$ to reach the launching point.
On the other hand, the size of the nozzle $A_*$ from Eq. (1) does not depend on $R_s$ and is an independent entity, which follows from observational data.

A further advancement in our model is the incorporation of gravitational effects, particularly relevant near compact objects. 
We modify the effective Mach number with a differential correction term for the local velocity to account for gravitational deceleration:
\begin{equation}
\mathrm{d}M(R) = -\frac{\mathrm{d}R}{c_{\rm s}}\frac{GM_{\rm src}}{R^2 \cdot V(R)}
\end{equation}
\noindent where $V_{\rm esc} (R)$ represents the local escape velocity and $M_{\rm src}$ is the mass of the  central object. 
The corrected lower value for $M(R)$ may then be used to recalibrate the flow properties, reflecting the gravity-induced velocity reduction. 
The (initial) Newtonian gravitational modification introduces a fundamental constraint on the subsonic flow: the requirement that the flow must achieve an escape velocity exceeding the local sound speed at the throat (launching point). This establishes a minimum temperature threshold that depends on the gravitational potential, determining the initial acceleration of the outflow.
 
By incorporating the above equations into a MatLab code, a model outflow shape can be generated based on the pressure gradient distribution parameters $\alpha$ and $R_{\rm s}$ and a distant $P_2$ where needed. 
Superposing this model shape on the observed shape allows us to constrain these values for each astrophysical outflow.  

\subsection{Outflow Simulations and Model Fitting} \label{sec:sec22}

The outflow evolution is governed by pressure balance at its boundary, where internal pressure arises from multiple contributions: kinetic pressure from viscous interactions between the high-velocity flow and ambient medium, magnetic pressure from entrained fields, and centrifugal pressure from rotational motion. These interactions create a structured boundary layer that shields the energetic core flow and produces observed features such as edge-brightening. While our current formalism focuses on the dominant pressure balance, future numerical simulations incorporating detailed boundary layer physics will be essential for understanding the complete evolutionary dynamics and environmental feedback mechanisms.

A Matlab code has been written to evaluate the above equations using a pressure gradient confinement in order to determine the flow parameters and the location of the boundary layer and the shape of the outflow.
Parameters for determining the shape of the outflow are the index $\alpha$ and the scale size $R_{\rm s}$ of the ambient pressure gradient and a variable describing the influence of gravity on the outflow close to its launching point.
The constant external $P_2$ pressure determines the shape in the outer part of the flow and causes the transition from diverging to a more parallel flow.

The physical parameters of a deLaval-like nozzle outflow vary systematically with distance as a result of the balance with the decreasing ambient pressure: the cross-sectional area increases and the density and temperature decrease, while the Mach number and the flow velocity increase. 
Figure \ref{fig:Mfig1} shows a simulation of an outflow subjected to a decreasing ambient pressure ($\alpha = -2$ and $R_{\rm s}$ = 10), where the Mach number increases to $M = 3.5$, while the flow velocity increases five fold across this range. 

\begin{figure*}
\centering
\includegraphics[width=0.49\columnwidth,angle=0]{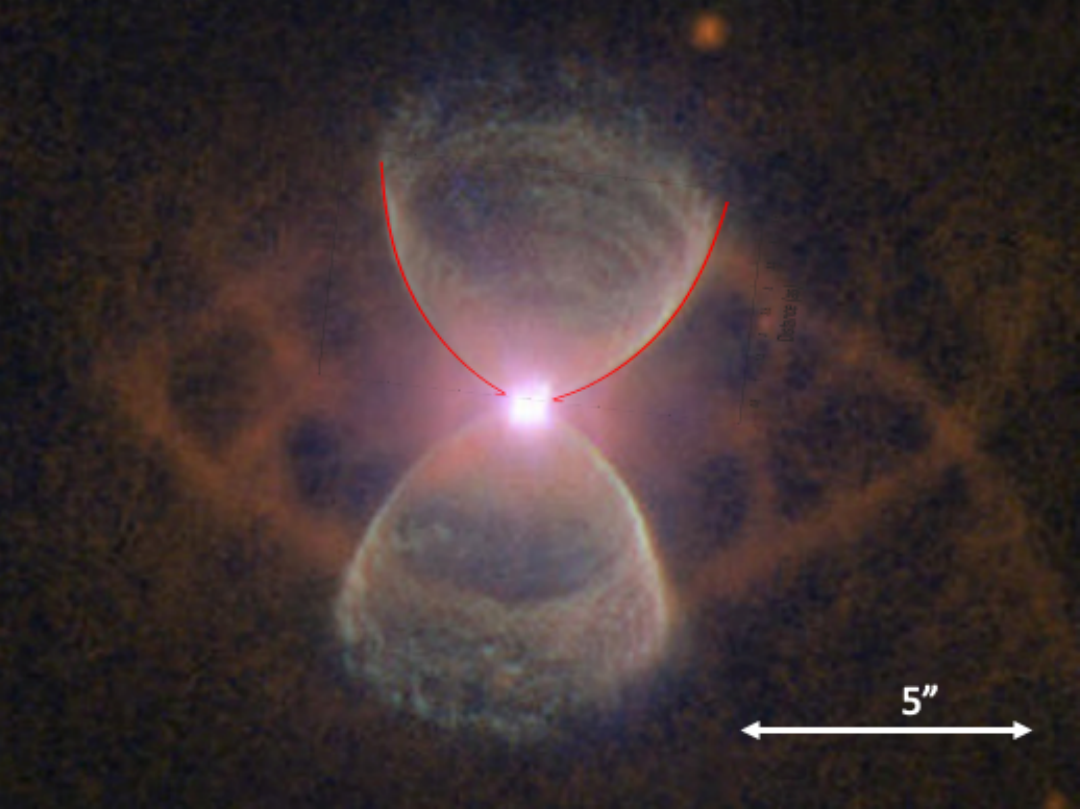}
\includegraphics[width=0.45\columnwidth,angle=0]{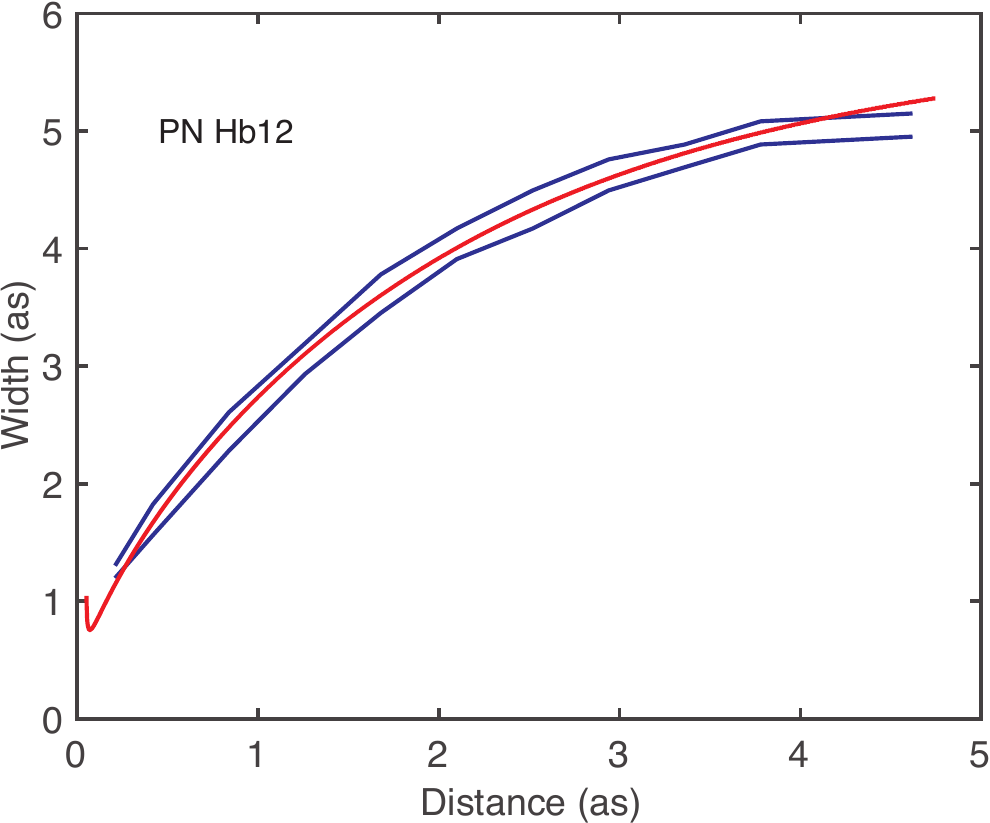}
\caption{Modeling the bipolar outflow in PN Hb\,12. 
(left) The model shape of the outflow superposed on a combined optical/infrared image of the inner structure of the Matryoshka Nebula Hubble Hb\,12.
This image with the Hubble Space Telescope using WFPC2 shows the bipolar Inner Nebula outflow and the Spiderweb shock structures of Hb\,12. 
(right) Model fitting of the very symmetric north and south outflow boundaries of the Nebula using a deLaval-analogy nozzle. 
The red model shape displayed is for a pressure gradient with index $\alpha = -2.05$, followed by a flat pressure at $6 \times 10^{-4}$ of the initial pressure. 
Photo: ESA/NASA/J. Schmidt.}
\label{fig:fig1}
\end{figure*}

While our current model focuses on the primary flow dynamics, detailed boundary layer evolution remains an area for future refinement
The very first flow parameter that defines any fluid boundary layer is the Reynolds number and its (possibly) magnetic variant.
This dimensionless number is defined as $R_{\rm e} = (V \cdot L)/\nu$, where $V$ is the flow velocity, $L$ is a scale length (or travel distance), and $\nu$ is a kinematic viscosity of the fluid.
At low flow velocities and low Reynolds numbers the flow pattern in the boundary layer will be laminar (sheet-like), while at high Reynolds numbers the flow at the interface becomes turbulent and starts to drain energy from and entrain material into the outflow.
Any such developments in the boundary will therefore determine the development of the flow pattern and the long-distance stability of the outflow.

A second simulation in Figure \ref{fig:Mfig1} shows the effect of gravitational deceleration on the outflow parameters. 
The throat of the outflow is moved along the pressure gradient after which the Mach number and the flow velocity increase rapidly, while also the flow area expands more rapidly making a wider opening angle.
The velocity increase will be reduced to a factor of three but the outer shape of the outflow remains the same.
This change during the initial expansion has been observed in the shape of the extragalactic sources, giving a more optimal fit and placing the nozzle closer to the nuclear source.

\begin{figure*}
\begin{center}
\includegraphics[width=0.48\textwidth,angle=0]{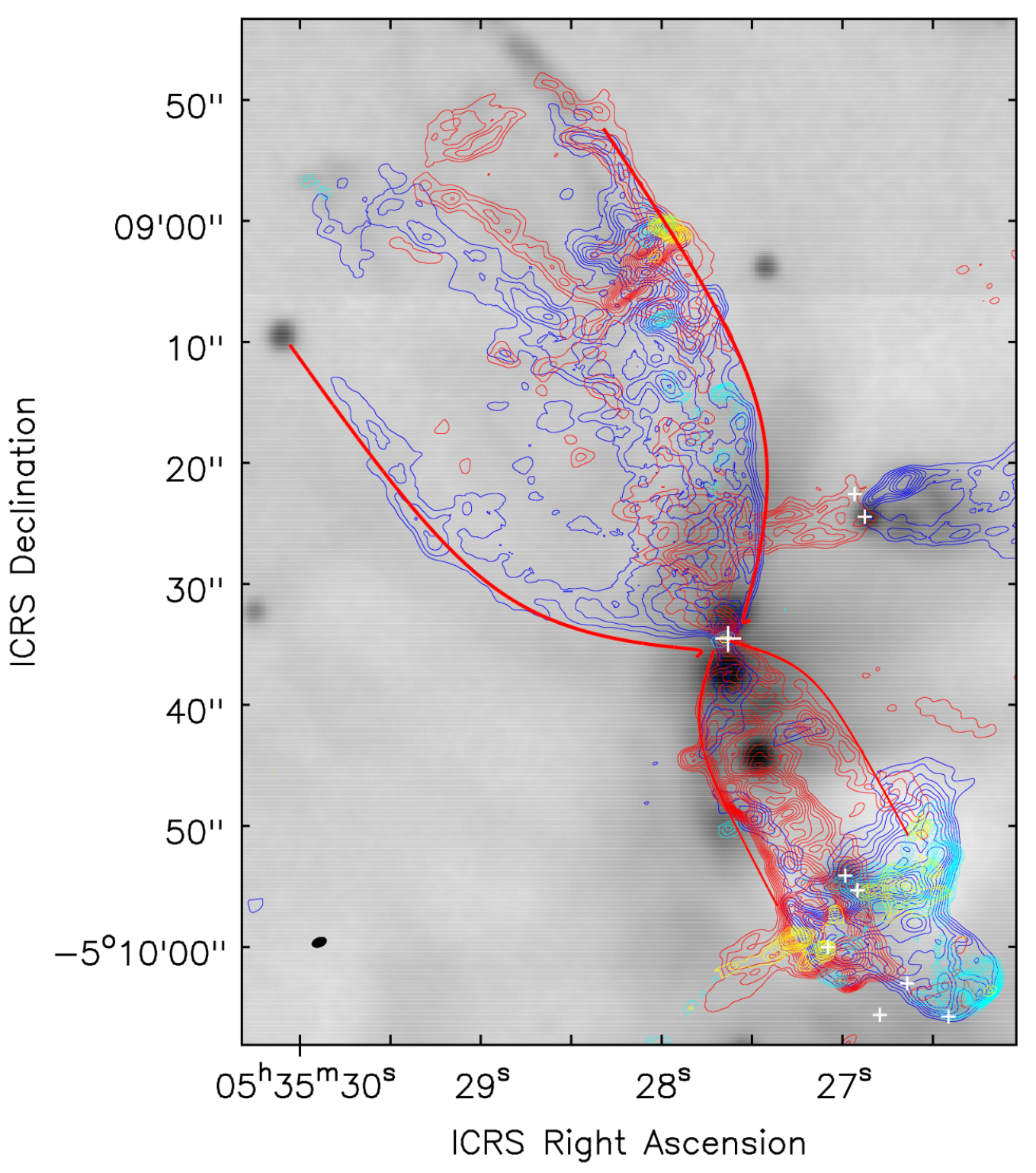}
\includegraphics[width=0.5\textwidth,angle=0]{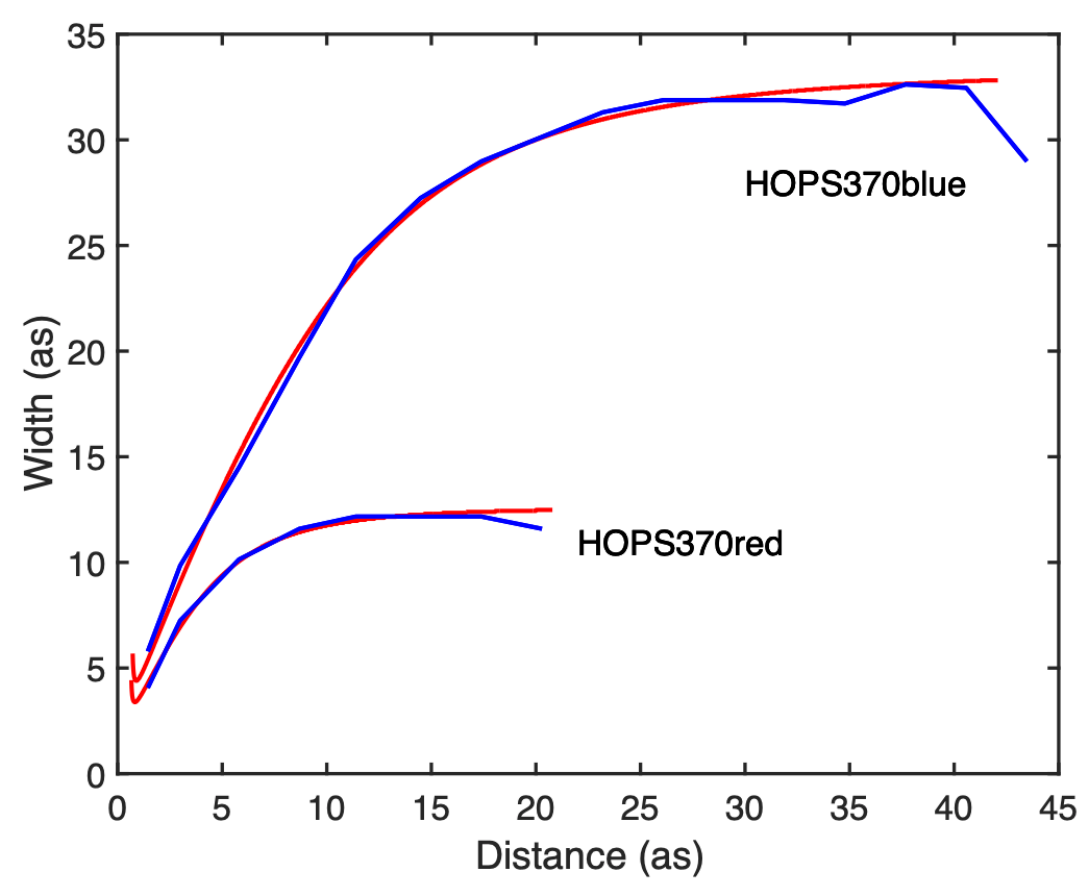}
\caption{Outflows of the Young Stellar Object HOPS 370 located in a field in Orion OMC-2 as presented by \citet{2023ApJ...944...92S}. (left) A model fit of the shape of the red and blue outflows is superposed on HOPS 370. (right) The observed widths (blue) and the deLaval nozzle model (red) for the two outflows fitted with an $\alpha = -2.8$. Although their sizes vary significantly and they show irregularities in the boundaries, their shapes resemble a standard nozzle outflow (1 as = 45 AU).}
\label{fig:fig2}
\end{center}
\end{figure*}

The measured outflow widths of the example sources have been used to constrain and optimize the values of $\alpha$ and $R_{\rm s}$ as well as to adjust a necessary value for $P_2$. 
The measured widths of the outflows together with the optimized fit are shown in Figures \ref{fig:fig1}-\ref{fig:fig4}.
The outflow shape associated with these optimized parameters was generated to be superposed on the images of the outflows as shown in these Figures.
A nominal gravity correction has been used only with the extragalactic sources, in order to show the effect on the shape of the outflow and to obtain a more precise fit with the observed shape at the launching region.
Gravitational corrections prove essential for extragalactic sources but appear negligible for Galactic objects, providing physical insights into the launching mechanisms across different scales.

\subsection{Data Sources}

For the planetary nebula Hb\,12, the Hubble Space Telescope optical emission line images have been used from \citet{2007ApJ...660..341K}. 
For the protostellar outflow HOPS\,370, Atacama Large Millimeter Array (ALMA) molecular line maps were used from \citet{2023ApJ...944...92S}. 
For the more distant FR\,I radio galaxy 3C\,84, the space-very-long-baseline-interferometry (SVLBI) continuum images at 22 GHz were adopted from \citet{2018NatAs...2..472G}. 
For nearby FR\, I M\,87, the image obtained with the global VLBI continuum observations at 86 GHz was used from \citet{2023Natur.616..686L}. 
See the references for full details on the observations and data reduction.

\section{Modeling Results}\label{sec:results}

The deLaval analogy model has been applied to well-defined outflows across diverse astronomical objects including YSOs, planetary nebulae (PN), and two extragalactic Fanaroff-Riley Type I (FR I) radio galaxies. 
To demonstrate the broad applicability and unification capability of the proposed paradigm, we showcase four representative examples in Figures \ref{fig:fig1} through \ref{fig:fig4}: the PN Hb 12, the YSO HOPS\,370 outflows, and the FR\,I jets in 3C\,84 and M\,87. 
Remarkably, quality model fits arise for all objects with the gradient exponent $\alpha$ consistently near minus two, pointing to ambient confinement as the potentially common governing factor in shaping astrophysical outflows. 
To illustrate our fitting results, we offer these four representative examples:

\begin{itemize}
    \item \textbf{Planetary Nebula Hb\,12.} 
The bipolar nebula Hb\,12 shows an hourglass-shaped outflow in optical emission (Fig.\ref{fig:fig1}(left)) \citep{2001ApJ...553..211C}. 
The northern and southern boundaries of the outflow can be fitted accurately as a function of distance (Fig. \ref{fig:fig1} (right) using an ambient pressure gradient exponent $\alpha = -2.05$ for the initial expansion combined with constant pressure at the outer regions of the outflow. 
The launching point lies within the extended stellar envelope at $\sim134$ astronomical units (AU), consistent with models for this source \citep{2014AJ....148...98C}.
The deLaval-like shaping in the low-power PN outflow Hb\,12 suggests that the collimation into an hourglass shape results from an ambient pressure gradient.

    \item \textbf{Young Stellar Object HOPS\,370.} 
The embedded Class I protostar HOPS\,370 (also known as OMC2-FIR3) in Orion drives a parsec-scale bipolar outflow extending out to 17,000 AU (Fig. \ref{fig:fig2}(left)) \citep{2017ApJ...840...36O, 2023ApJ...944...92S}. 
The blue lobe exhibits the characteristic deLaval shape with a steeper pressure gradient with $\alpha = -2.8$ (Fig. \ref{fig:fig2}(right)) for the initial expansion and a lower pressure for its outer region. 
The estimated launching point is $\sim330$ AU, consistent with expectations for an accretion disk origin. 
The red lobe of HOPS-370 has a shape with a similar  $\alpha = -2.7$, but encounters a higher density ambient medium that also causes deformation of the outflow.
Both the blue and red outflows show an initially conical outflow followed by a cylindrical shape because of a fixed external pressure in the outer regions.
The outflows of HOPS\,370 extend the deLaval analogy to massive YSOs and show the collimation of powerful outflows from forming stars over sub-parsec scales.
    
    \item \textbf{FR\,I Radio Galaxy 3C\,84 - Centaurus A.} 
A model fit of the shape superposed on the edge-brightened sub-parsec launching area of the one parsec-scale radio jet in 3C\,84 \citep{1990MNRAS.246..477P,2018NatAs...2..472G}  (Fig. \ref{fig:fig3}(left)). 
The conical shape defines the boundary layer of the flow and there is weak evidence of the central high-power jet component.
This gravity-corrected model reproduces accurately the observed width of the inner regions of the jet outflow with $\alpha = -2.2$, while the outer regions become more cylindrical with a flat pressure profile (Fig. \ref{fig:fig3}(right)).
Flow instabilities start already at a distance of 0.35 parsec (1 mas) from the core.
The effect of gravity close to the nucleus gives an improved fit to the initial expansion region and places the launching region at 33 gravitational radii $R_{\rm g}$ inside the ergosphere of the supermassive black hole (SMBH) with a mass of $2 \times 10^9 \Msol$. 
Pressure-gradient shaping of the outflow works for this low-power relativistic extragalactic jet launched near the black hole. 
    
\begin{figure*}
\begin{center}
\includegraphics[width=0.5\columnwidth,angle=0]{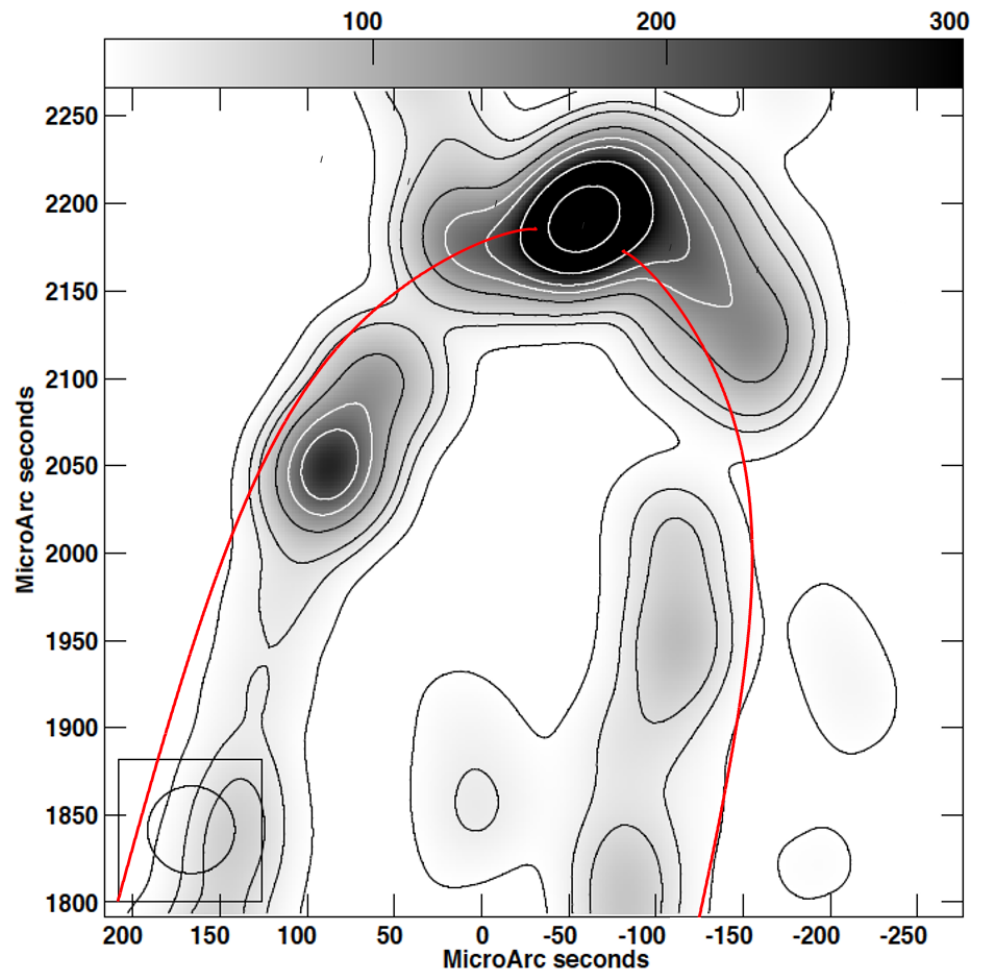}
\includegraphics[width=0.49\columnwidth,angle=0]{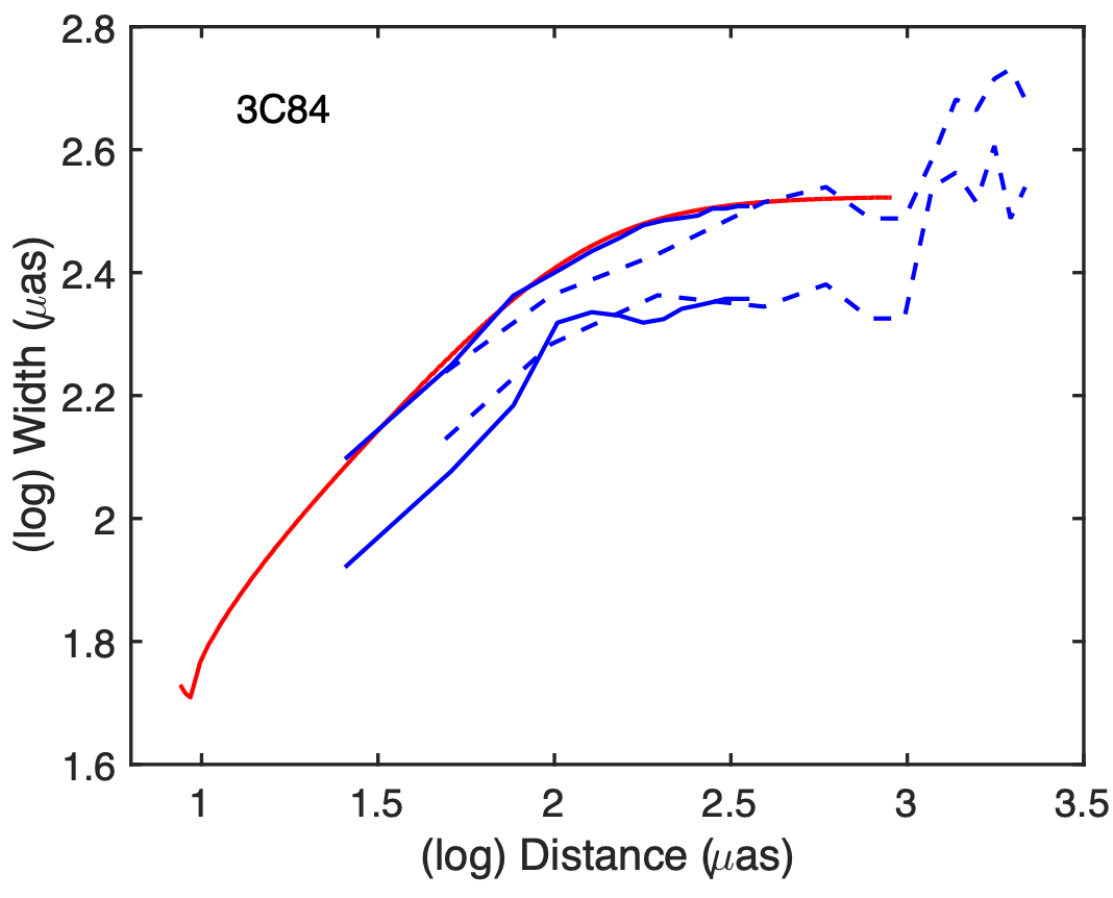}
\caption{The outflow shape of the low-power jet in FR\,I 3C\,84 showing the edge-brightened boundary layer and some evidence of the high-power core-flow. 
(left) The model shape is superposed on the outer regions of the boundary layer of the jet outflow.
(right) The width of the deLaval model with gravity compared with the observed width of the outflow.
The upper curves represent the outer contours of the edge-brightened jet structure and the lower curves follow the emission peaks.
The solid curves have been taken from the high-resolution image (left) and the dashed curves from a lower-resolution image in \citet{2018NatAs...2..472G}.
The red curve represents an outflow shape in a pressure gradient with $\alpha = -2.4$, followed by a flat ambient pressure in the outer region at a factor $1.9 \times 10^{-3}$ below the nozzle pressure.
The model shape shows the onset of the fixed ambient pressure starting at 120 $\mu$as (430 $R_{\rm g}$ = 0.041 pc) and the jet starts to oscillate at 630 $\mu$as (2,260 $R_{\rm g}$ = 0.22 pc).
The launching point for this case is located at 32 $R_{\rm g}$ (1 mas = 0.344 pc = $3.58 \times 10^3 \, R_{\rm g}$). }
\label{fig:fig3}
\end{center}
\end{figure*}
    
    \item \textbf{FR\,I Radio Galaxy M\,87.} 
The Messier galaxy M\,87 hosts a prominent jet 1.5 kpc in length \citep{1989ApJ...340..698O} that feeds into a very extended radio lobe structure.
The structure of the jet launching region is defined by the edge-brightened shape of the boundary layer and the central high-power jet component.
A gravity-corrected fit to the shape of the launching area of the jet is superposed on a high-resolution map of the outflow near the nucleus (Fig. \ref{fig:fig4}(top)) \citep{ 2023Natur.616..686L}).
The conical jet geometry in this inner region (up to 0.65 pc) can be modeled very well with gravity included and with $\alpha = -1.95$, but without any constant pressure contribution for the fitted range of the outflow (Fig. \ref{fig:fig4}(bottom)). 
For comparison, a non-gravity-corrected fit of the outflow has been added in the graph.
The jet remains conical beyond the region considered here until instabilities set in at 1.23 kpc, at which point the jet becomes cylindrical.
The launching point in the inner collimation region at 13 $R_{\rm g}$ fits with the ring-like accretion structure at the nucleus of the $6.5 \times 10^9 \Msol$ SMBH and the outer radius of the resolved nuclear disk. 
The M\,87 jet extends the applicability of the model and connects the jet origin to the SMBH environment.
\end{itemize}

A striking feature of our results is the narrow range of pressure gradient indices ($\alpha \approx 2.0$–$2.8$) required to fit outflows across six orders of magnitude in spatial scale. 
The fitting parameters for these four examples are presented in Table \ref{tab1}.
This consistency suggests that ambient pressure confinement represents a fundamental mechanism in outflow physics, transcending the specific details of the central engine or the launching mechanism. 
Furthermore, the systematic variation in $R_s$ values correlates well with the characteristic scales of the host systems, providing an independent validation of our model's physical basis. 
The gravitational modification proves particularly important for extragalactic sources, enabling more accurate fits near the central engine while maintaining consistency with observed large-scale structures. 
This refinement addresses a long-standing challenge in jet modeling by naturally explaining both the initial wide-angle expansion and subsequent collimation without requiring additional mechanisms.

\section{Discussion}

Our outflow analysis reveals a fundamental universality in astrophysical outflows, despite their diverse origins, physical scales and environments. 
The morphological structure and geometrical evolution of all cases show similar environmental collimation, suggesting that the shapes of diverse astrophysical outflows can be unified using a deLaval-like nozzle analogy defined by appropriate pressure gradients of the ambient medium. 
Our model is based on key assumptions that allow us to focus on the fundamental physics of outflow shaping: the outflow is treated as a pressure-confined supersonic flow; the ambient medium provides a continuous pressure gradient; the boundary layer maintains pressure balance with the ambient medium. While factors such as rotation, magnetic fields, and relativistic effects may play important roles in outflow physics, we suggest and demonstrate that the pressure gradient dominates the large-scale morphology. This simplified approach across different scales suggests that pressure confinement represents a fundamental mechanism in outflow collimation.
The model parameters for the above example are presented in Table \ref{tab1}.
The launching regions are consistently near the central engine, whether a stellar envelope, an accretion disk, or the vicinity of a SMBH.
Outflows expand initially conically in a steep pressure gradient before slowly transitioning to a more cylindrical structure once entering regions with more constant pressure at longer distances or because of changes in the structure of the boundary layer.

The modeling examples also reveal systematic differences between the Galactic outflows and the extragalactic examples.
The launching points for the Galactic cases appear relatively far away from the source where gravity only marginally impacts the outflows.
As a result, the low-power outflows of PN Hb\,12 and YSO HOPS\,370 exceed significantly the minimum launching requirements such that the outflow velocities predicted by our models are consistent with observed velocities (see Section below).
On the other hand, for the extragalactic sources, the models determine the location of the launching points (the nozzle or the base of the jet) at tens of gravitational radii $R_{\rm g}$ from the nucleus, which suggests very high launching temperatures and relativistic flow velocities. 

\begin{figure*}
\begin{center}
\includegraphics[width=0.7\columnwidth,angle=0]{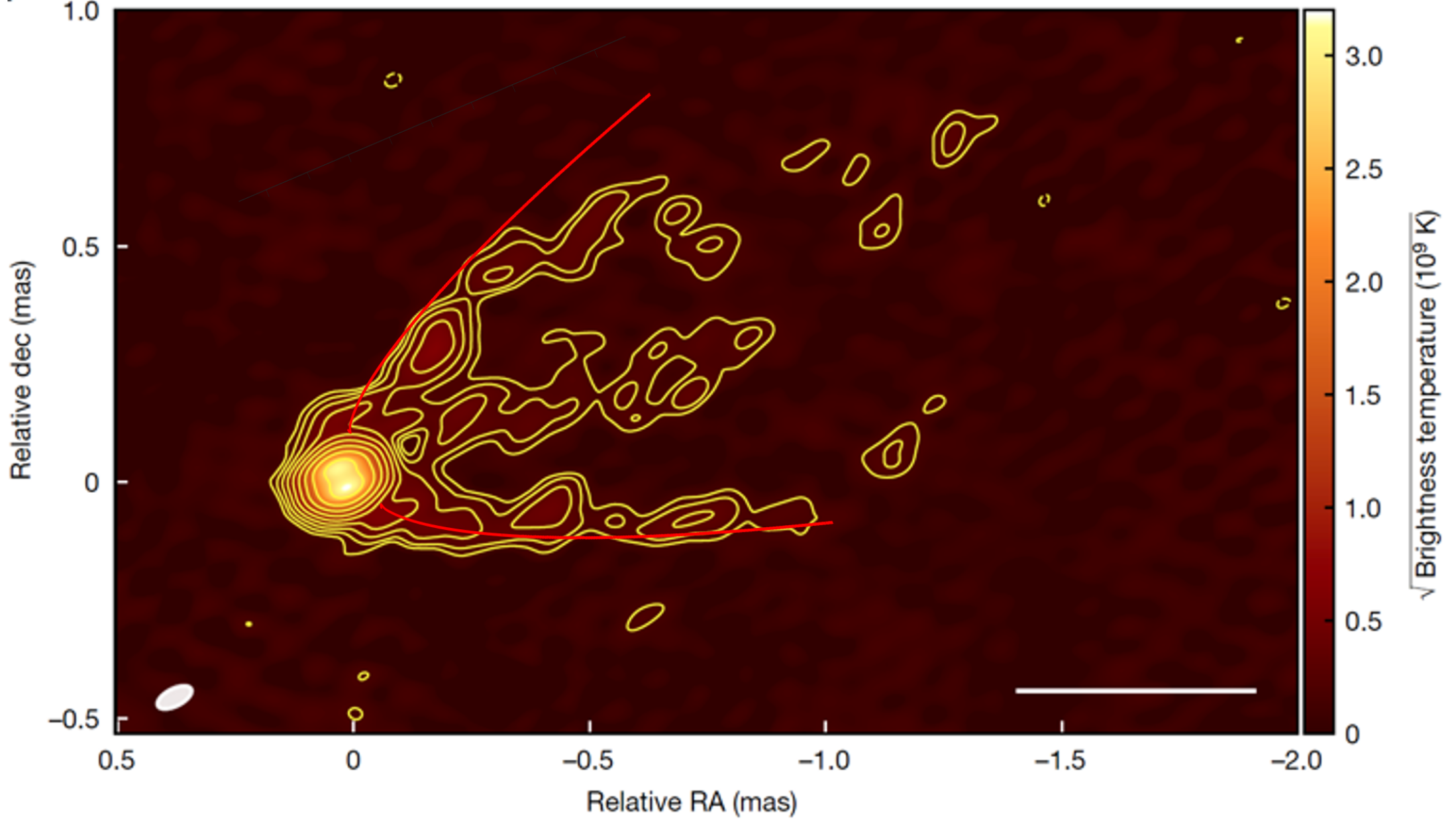}
\includegraphics[width=0.6\columnwidth,angle=0]{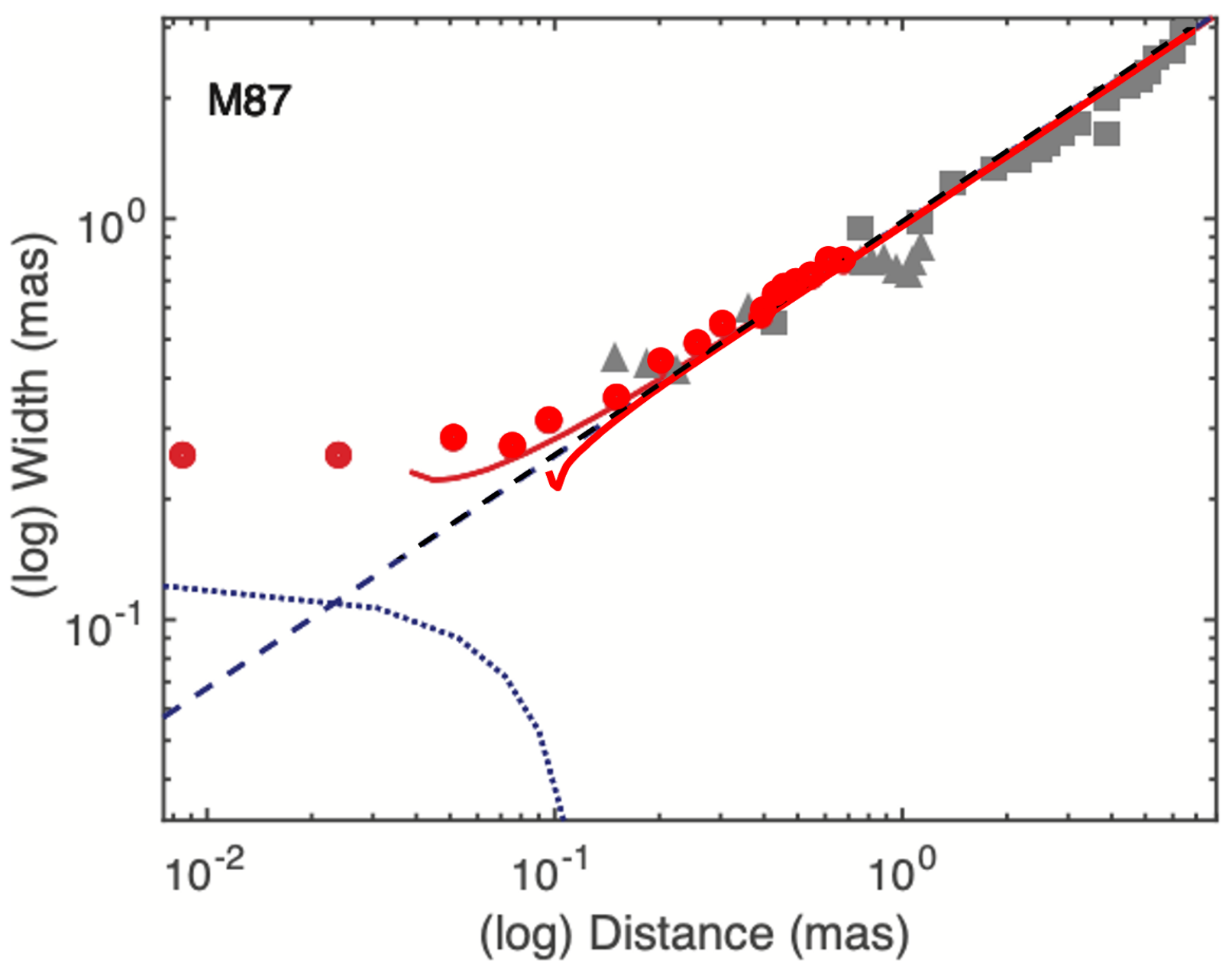}
\caption{The shape of the launching region of the outflow/jet in FR\,I M\,87 - NGC\,1275 showing the edge-brightened boundary layer and the high-power core flow.
(top) The model shape superposed on the higher resolution map of the inner regions of the jet in M\,87 as presented by \citet{2023Natur.616..686L}.
(bottom) The model fit of the width of the observed jet using a pressure gradient with $\alpha = -1.95$, where the upper red curve is without gravity effects and the bottom red curve is with gravity.
The red and grey data points have been adopted from the high-resolution data and referenced papers as presented in \citet{2023Natur.616..686L}).
The model shape extends along the whole jet until it becomes unstable at 1.4 kpc.
The (left bottom) dotted curve represents twice the diameter of the nuclear ring of 8.4 $R_{\rm g}$ (or 64 $\mu$as) found in the nucleus \citep{2023Natur.616..686L}.
The launching point of the nozzle in the model is at 13 $R_{\rm g}$ and lies close to the outer edge of the ring structure.
(1 mas = 0.0814 pc = 131 $R_{\rm g}$)}
\label{fig:fig4}
\end{center}
\end{figure*}

\subsection{The Boundary Layer}

An important aspect of outflow modeling has not yet been incorporated in the simulations, but has been proven to be crucial for jet evolution: the boundary layer between the fast-moving outflow and the ambient medium.
This boundary layer maintains pressure balance with the ambient medium and serves as a protective shield surrounding the energetic flow at the center of the outflow.
In a high Reynolds number scenario, the viscous interaction in the boundary layer adjusts the mutual pressure balance with the external medium and modifies the transverse velocity profile (see Section \ref{sec:sec22}). 
The resulting energy dissipation and momentum transfer in this protective layer account for the observed edge-brightening of the outflows. 

Numerical studies including viscosity show that the detailed physics within the outflow boundary layer causes downstream morphological changes \citep{1980ApJ...239..433B,1997MNRAS.286..215K,2009MNRAS.397.1113W}. 
As the outflows move away from the throat launching point, the boundary layer thickens rapidly and the transverse velocity profile varies systematically with distance.  
Different viscosity mechanisms imprint distinct transverse velocity profiles and emission patterns on the outflow, potentially offering diagnostic tools for identifying the primary physical processes. Comparison of simulated emission patterns with multi-band observations may thus elucidate the particular processes, whether Coulomb collisions, turbulence, plasma interactions, or magnetic field effects govern the momentum and energy transport across the boundary layer \citep{1980ApJ...239..433B, RicciEA2024}. 

The characteristics of the boundary layer also clarify the re-collimation phenomenon observed in low-power outflows. 
While the expanding boundary in a low-power source allows isentropic lateral expansion \citep{2009MNRAS.397.1113W}, the strongest viscous dissipation gradually moves inwards and narrows the high-velocity core with distance, which causes re-collimation at longer distances \citep{1980ApJ...239..433B}. 
In more relativistic jets, instead, the boundary layer maintains the overall pressure balance while preserving and protecting the high momentum flow in the centre. 
Tracking this boundary layer evolution in different sources will reveal the confining role of pressure gradients. 

Additionally, shear-induced turbulence in the boundary layer may trigger multiple instability modes that can disrupt FR I jets \citep{1981MNRAS.196.1051F,1982ApJ...257..509H}. 
These include Kelvin-Helmholtz instabilities at the jet-ambient medium interface and current-driven instabilities in magnetized regions. As jets decelerate, surface instabilities become increasingly dominant, creating edge-brightened features and filamentary structures when velocities fall below some critical threshold. 
The growth and saturation of these instabilities depend on local jet conditions including the velocity profile, density contrast, and magnetic field configuration, ultimately leading to the termination of collimated beams characteristic of FR I jets \citep{2012ApJ...760...77A}. 
While some studies suggest jet precession causes the oscillatory patterns \citep{2023Natur.621..711C}, the observed morphology can be naturally explained by the interplay between these instabilities and magnetic flux eruptions that drive large-scale surface waves \citep{ChatterjeeEA2023, NikonovEA2023}.

\begin{table}
\centering
\caption{Modeling parameters of astrophysical outflows}
\begin{tabular}{lccccccc}
\hline
Source & Alpha & Scale  & Level    & Distance & Size   & Min Launch  & Extent \\ 
Name &         & Height & Pressure & Throat   & Throat & Temp & Outflow \\
\hline
PN Hb\,12 & $-2.05$ & 107 AU & $6\times 10^{-4}$ & 134 AU & 1630 AU & 960 K& 10060 AU \\
HOPS\,370blue & $-2.8$ & 280 AU & $1.9\times 10^{-4}$ & 330 AU & 1750 AU & 1035 K& 16670 AU \\ 
HOPS\,370red  & $-2.7$ & 250 AU & $2.1\times 10^{-3}$ & 330 AU & 1520 AU & 1035 K& 7940 AU \\
FR\,I 3C\,84 & $-2.2$ & 12 $R_{\rm g}$ & $1.8\times 10^{-3}$ & 32 $R_{\rm g}$ & 207 $R_{\rm g}$ & $4\times 10^{11}$ K & 0.35 pc \\
FR\,I M\,87 & $-1.95$ & 3 $R_{\rm g}$ & -- & 13 $R_{\rm g}$ & 26 $R_{\rm g}$ & $5\times 10^{11}$ K& 1230 pc \\
\hline
\end{tabular}\\
Note 1: The assumed distance for Hb\,12 is 2.24 kpc, but it  is likely to be smaller.
Note 2: The distance of HOPS\,370 in the Orion Molecular cloud is 398 pc.
Note 3: For 3C\,84, 1 mas = 0.344 pc = $3.58 \times 10^3 R_{\rm g}$.
Note 4: For M\,87, 1 mas = 0.0814 pc = $131 R_{\rm g}$.
Note 5: The minimum launching temperature for HOPS\,370 and Hb\,12 is based on a mass of 2.3 \Msol\ and 1 \Msol, respectively.
\label{tab1}
\end{table}

\subsection{Launching the Outflow}

The large-scale phenomenology associated with this pressure-confined fluid outflow scenario appears largely consistent with observations, while no details yet exist yet about what could happen before the launching point of the supersonic flow.
In order to get outflow from the nozzle, there needs to be a supply of material in the atmosphere at a high enough temperature for the sound speed to exceed the escape velocity. 
An 'overpressure' in the lower atmosphere at the polar region may then drive some convection-like flow preferably along the symmetry axis of the system, the path of the steepest descent.
Material in the high pressure atmosphere surrounding an accreting source has been viscously preheated in the disk and possibly magnetically heated in the inner section of the disk, and may experience further heating close to a compact source. 
In the atmosphere, the heated material will experience the decreasing pressure gradient, which would start subsonic acceleration outwards. Upon reaching the local sound speed, the flow will continue supersonically and form a distinct outflow or jet along the symmetry axis of the source.
In this process, the flow will carry along any embedded rotational component and will entrain locally embedded magnetic fields as may also be observed in Solar prominences.
The more the local sound speed exceeds the escape velocity, the higher will be the power of the outflow.


The outflow does act as a 'release valve' for the lower regions of an overheated atmosphere and  the observed supersonic expansion after the launching point does suggest that there must also be an underlying subsonic flow in the atmosphere of the source.
The requirement in the deLaval nozzle scenario that the sound speed at the launching point is (much) larger than the escape velocity determines a required minimum local temperature and velocity for launching the outflow.
The minimum launching temperatures for the five fitted outflow sources have been presented in Table \ref{tab1};
while in reality, the actual launching temperatures for the observed outflows will be significantly higher. 

\subsection{Comparing Galactic Outflows}

The existence of pressure gradients surrounding lower-power outflows provides a satisfactory explanation of their shape as observed in the Galaxy. 
Well-defined accretion driven outflows in Planetary Nebulae as found in Hb\,12 (Fig. \ref{fig:fig1}) and several other PNe (not presented here) can be fitted with pressure gradients with an index close to 'minus two'.
Early modeling of such sources have assumed a conical outflow, which may give reliable flow parameters but does not provide a correct shape \citep{2015A&A...582A..60C}.
Incidentally a conical outflow is observed in PN Red Rectangle (see \citet{2016A&A...593A..92B}), but here the nozzle outflow may be explained with a steep gradient with index of 'minus 3.2'. 

The low-power outflows observed in the YSO HOPS\,370 appear to be representative for most or all YSO sources.
The above fitting results indicate that these source may be subject to steeper pressure gradients existing in these star formation regions. 
However, no other modeling results have yet been found in the literature for making comparisons.

The simplified deLaval modeling presented in this work already provide a first handle for determining some representative flow parameters for well documented low-power sources. 
For the Galactic source Hb\,12 the minimum launching velocity translates to 3.6 \kmss, but with a measured surface temperature of 35,000~K, the actual sound speed velocity would be 22 \kmss. 
Together with nominal supersonic acceleration in the outflow of a factor of five (Fig. \ref{fig:Mfig1}), the Hb\,12 flow velocity would be on the order of 110 \kmss, to be compared with observed flow velocities in the boundary of 90 \kms \citep{2014AJ....148...98C}.
The size of Hb\,12 is on the order of 1\arcsec\ or 2,245~AU, if it is located at a distance of 2.24 kpc, which appears consistent with the scale size deduced from the fitting in Figure \ref{fig:fig1}.

Similarly for HOPS\,370 with an estimated mass of 2.3 \Msol, the minimum launching velocity is 3.5 \kmss. 
For a surface temperature of the star similar to a main sequence star at 11,000K, the actual launch velocity at the nozzle would be 11.4 \kmss. 
With a velocity acceleration of a factor of 5, this would produce a velocity of about 57 \kmss, which is again comparable with the observed velocity of 51 \kmss\ \citep{2023ApJ...944...92S}.
The estimated size of HOPS\,370 of 499 AU is also comparable with the fitted parameters. 


\subsection{Comparing ExtraGalactic Outflows}

The characteristic shape of a low-power wind/outflow has also been observed in the nearby galaxy NGC\,3079 \citep{cecilEA2001}.
This large-scale outflow above the plane of the disk at the location of the nucleus may have resulted from intense starburst activity in the nuclear region, but its shape can be well-fitted with the same 'minus two' index as for PN Hb\,12.

Modeling efforts of high-power jets as found in extragalactic sources have mainly focused on the high-momentum core of the jets without considering the interaction with an ambient medium, see \citet{Marti2019} for a review.
Such modeling efforts may adequately describe the conical/cylindrical shape of a jet away from the launching point assuming a (nearly) flat pressure profile and viscous internal expansion.

As explained in previous sections, the high power and high momentum components in powerful radio jets are contained in the center of the jet and the supersonically expanding boundary layer serves as a protective cocoon that balances the interaction with the ambient medium.

These modeling efforts do not automatically describe the parabolic expansion near the launching point. 
A recent model by \citet{MartiEA2016} has invoked viscosity effects to explain this expansion, which is similar to our early study of the jet in NGC\,315 from 1980. 
A next version of the model would introduce a pressure gradient.
A new study of the jet in NGC\,315 explains the initial parabolic expansion by introducing a thermal energy component in the flow, without describing the nature of this component \citep{RicciEA2024}.

In reality, the observed parabolic expansion close to the launching point confirms a supersonic flow into an ambient medium with a pressure gradient that forces the supersonic expansion of the viscous boundary layer that surrounds the high momentum flow observed in the center of the flows in both 3C\.84 and M\,87.


\subsection{The Effect of Gravity on the Outflows}

Gravitational retardation is shown to be significant for shaping the high-power FR\,I jets of the extragalactic SMBH sources.
Gravity appears to play no significant role for low-power sources because the launching points appear to be located away from the sources.
In this work, the introduction of a Newtonian deceleration term has been aimed at showing the effect of gravity on the shape of the supersonic flow and has not yet been used to optimize this shape with regard to the observed launching shape of the SMBH jets. 

The launching points follow from fitting the shape of the supersonic regions of the outflow, these a priori unknown points are close to the central source, and more appropriate general relativity arguments may be introduced to describe both the subsonic and supersonic regions of the jet outflow. 
Indeed, other simulations have addressed this issue by introducing an effective gravitational potential \citep{PaczynskyW1980} to include rotation \citep{DihingiaEA2018} in the outflows from a Kerr black hole.
However, these aspects are not easily introduced into our simplified model and require more complete numerical simulations.

\subsection{Future Observational Verification}

Our pressure-gradient model makes several testable predictions that future observations could validate and refine. High-resolution VLBI observations at frequencies above 22 GHz would be particularly valuable for resolving the critical transition region between subsonic and supersonic flow. Simultaneous multi-frequency observations could map the spectral index distribution near the launching point, testing our predictions about temperature and velocity gradients. The Event Horizon Telescope and future enhanced VLBI arrays could directly image the nozzle region in nearby sources like M87, providing unprecedented views of the launching mechanism.

The model predictions about boundary layer physics could be tested through deep radio imaging designed to detect the predicted edge-brightening effects. Multi-wavelength observations combining radio, X-ray, and optical data would trace boundary layer evolution and energy dissipation processes. Polarization measurements could reveal how magnetic field configurations in the boundary layer relate to the pressure gradient, providing insight into the interplay between magnetic and pressure forces.

Environmental factors play a crucial role in our model, and their effects could be verified through X-ray observations measuring ambient medium properties along jet paths. Statistical studies comparing outflow shapes with host galaxy properties could reveal how environmental conditions influence pressure-gradients. The relationship between pressure-gradient indices and local ISM/ICM conditions would be particularly informative for understanding the universality of our model across different scales.

Time-domain studies offer another avenue for testing the model. Long-term monitoring of young stellar outflows could track shape evolution, while studies of multiple outflow episodes in single sources would test the consistency of pressure confinement mechanisms. Observations of disrupted and restarted jets could illuminate how pressure gradients affect jet stability and reformation.

These observational tests would benefit significantly from next-generation facilities like the ngVLA and SKA, particularly when combined with X-ray and optical observations. Such multi-wavelength, high-resolution data would provide comprehensive tests of our model predictions about how pressure gradients universally shape astrophysical outflows. The combination of these observations with improved numerical simulations would allow for rigorous testing of the pressure-gradient confinement mechanism across the full range of astronomical scales.

\section{Conclusions}

In a simplified manner, a pressure confinement model provides the analytical framework for unifying key outflow physics for compact objects through basic fluid dynamical principles. 
The success of the model in reproducing the observed supersonic morphologies confirms the presence of an expected pressure gradient in the medium surrounding compact sources and demonstrates that the subsonic part of the outflow originates inside the extended atmosphere of the (compact) source.
While the atmosphere around the compact source would be maintained by heated material from direct accretion and from indirect winds from a disk, 
the ambient pressure in the atmosphere at the viscous boundary layer interface is balanced by pressure resulting from dissipation and from thermal, rotational, and magnetic contributions.

The shape of the observed outflows confirm supersonic flow as predicted in a deLaval nozzle, because if it were not so, the outflow would just form a blob outside the atmosphere.
Particularly significant is our finding that the boundary layer between flow and ambient medium would behave as a fluid system and maintain the pressure balance largely independent of internal flow conditions, enabling this pressure confinement formalism to predict both the shape of the outflows and their subsequent evolution.

For low-power sources in the Galaxy, these results provide a way for determining the first order properties of the outflows by comparison with observations.
For extragalactic jets, detailed observations are still limited and jet properties may follow from assumed flow characteristics and representative fluid simulations.
Viscous fluid simulation may further describe the outflow properties and the behavior close to the launching point similar to our earlier viscous fluid simulations that described the shape of the jet and the development of the boundary layer.

The deLaval scenario clearly places the launching point inside the atmosphere at the polar region close to the source and this may work in a manner that is similar for both stellar sources and SMBH sources.
For extragalactic sources, this scenario may prefer  the presence of a Kerr SMBH, which has a clear symmetry axis and a high-pressure extended atmosphere surrounding the ergosphere. 
The relativistic region of powerful jets with embedded magnetic fields and rotation is concentrated/hidden in the center of the flow as it leaves the launching point and would not yet directly be affected by the supersonic expansion of the initial boundary layer as it interacts with the ambient medium.

The viscous boundary layer will drain energy and momentum from the (rotating/magnetized/relativistic) fluid at the core of the outflow, which will affect the large-scale performance and stability of the outflow and its travel distance.
Therefore, numerical simulations of both low- and high-power outflows can only be complete and realistic when addressing the fluid characteristics of the boundary layer and the ambient pressure profile along the path. 
Rigorously testing against both simulations and multi-wavelength data across diverse objects will further refine the scenario of shaping astrophysical jets by ambient pressure confinement.
In addition, the introduction of a simplified form of gravitational deceleration into our scenario does affect the shape of the outflows in powerful SMBH sources, which suggests that further numerical simulations also need to include general relativity at the launch of the outflow.

During the course of this project, the authors became aware of the multiple applications where this outflow scenario may apply: bipolar PN, Be stars and symbiotic star systems, Young Stellar Objects in star formation regions and Water Fountains, microquasars and possible accreting neutron star systems, star-formation driven galactic outflows as observed in nearby NGC\,3079, and of course extragalactic nuclear outflows and jets.
In addition, the scenario may also apply to the generation of the Solar wind.
This universality highlights the fundamental nature of pressure-confined outflow physics in shaping diverse astrophysical phenomena.

While our simplified model does not yet capture all the physical processes involved in the launching and propagation of the outflow, its success in describing outflow morphologies in diverse sources suggests that pressure confinement plays a fundamental role in astrophysical outflow physics. Future work combining this pressure-gradient framework with detailed treatments of rotation, magnetic fields, and relativistic effects will provide a more complete understanding of outflow dynamics.

\section{Acknowledgments}
This work is partly supported by National SKA Program of China (2022SKA0120102) and National Natural Science Foundation of China (No. 12041301). 
WAB and TA acknowledge the support from the Xinjiang Tianchi Talent Program.

\bibliographystyle{aasjournal}
\bibliography{sn-bibliography}

\end{document}